\documentstyle[11pt,newpasp,twoside,epsf]{article}
\def\edcomment#1{\iffalse\marginpar{\raggedright\sl#1\/}\else\relax\fi}
\reversemarginpar

\begin{document}
\title{Truncations in stellar disks}
 \author{P.C. van der Kruit}
\affil{Kapteyn Astronomical Institute, University of Groningen,\\
P.O. Box 800, 9700 AV Groningen, the Netherlands\\
email: vdkruit@astro.rug.nl}

\section{Background.}

Stellar disks in edge-on spiral galaxies have been known to have --at
least in some cases-- rather sharp edges or truncations (van der Kruit
1979). These are obvious when examining isophote maps in which 
the outer isophotes in the radial direction come suddenly much closer 
together (see Fig. 1). The reality of the truncations can also be 
inferred from the fact that the
fainter isophotes of bright stars and along directions perpendicular to
the galaxy plane in Fig.1 do not show the decrease in spacing.

Truncations can also be inferred from the fact
that often diameters of edge-on galaxies appear not to be growing
between early sky surveys (using IIa-emulsions) and later,
deeper surveys (IIIa-emulsions). Bosma \&\ Freeman (1993) 
compared diameters of galaxies on the {\it Palomar Observatory
Sky Survey} prints
with those on the {\it SRC-J Survey}. About a quarter of the
galaxies showed no significant increase in diameter, suggesting a cutoff
in the outer disks. They calibrated their SRC limiting
isophotes as about 25.5 B-mag arcsec$^{-2}$, which is at about the level
where often the truncations become visible in edge-on galaxies (but not in
more moderately inclined ones).

Van der Kruit \&\ Searle (1981a,b, 1982) found in a sample of 7 edge-on
spirals that the truncations occured at a galactocentric
distance of 4.2 $\pm $ 0.5 exponential scalelengths $h_{\rm R}$ of
the surface brightness distribution. 
Many moderately inclined spiral galaxies have a (face-on) central surface
brightness of about 21.7 B-mag arcsec$^{-2}$ (Freeman 1970). Considering
that the limiting surface brightness in many photometric studies is
about 26 to 27 B-mag arcsec$^{-2}$ and that the truncations occur at 
4 or 5 radial scalelengths, it would be very difficult to see the
cutoffs in face-on galaxies. Furthermore, it would be misleading to
look for these in published
radial surface brightness profiles, since these are usually produced by
azimuthally averaging the observed surface brightness maps; this
procedure smoothes out truncations when these are not exactly circular. 

Van der Kruit (1988) examined the isophote maps of the Wevers et al.
(1986) sample and found that in most systems the outer contours were
more closely spaced than the inner ones, suggesting a drop in
scalelength by at least a factor two (see Fig. 2). 
Four systems did not show this effect and the disks seemed to extend 
out to 6 or 7 scalelengths; interestingly, 
these were all of early type. The remaining 16 systems did show
the effect and this suggested a truncation at $R_{\rm max} = 4.5 \pm 1.0\;\;
h_{\rm R}$. Isophote maps of Roelof de
Jong's sample (de Jong \&\ van der Kruit 1994) show the effect also 
in many cases.
\begin{figure}
\plotone{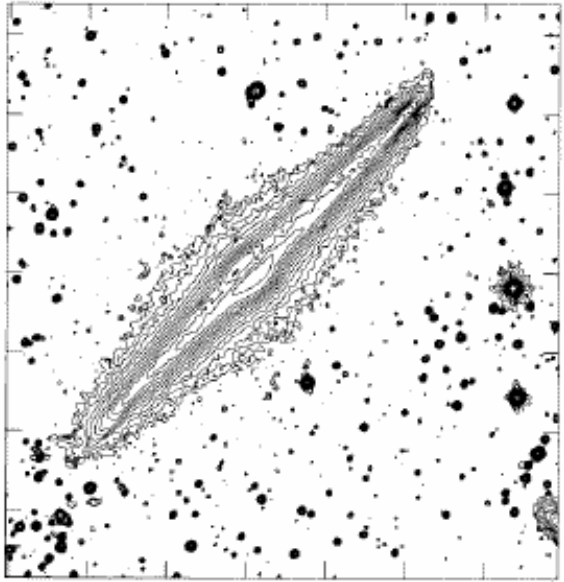}
\caption{Isophote map of the edge-on spiral NGC 4565 (van der Kruit 
\&\ Searle 1981a). The isophote interval is 0.5 mag. Note the sharp
decrease in surface brightness at the ends of the disk, while along the
vertical directions and for the bright stars there is no such steeper decline.}
\end{figure}

\begin{figure}
\plotone{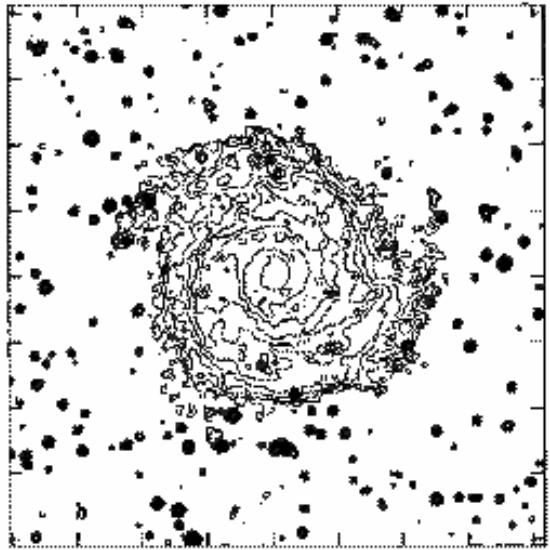}
\caption{Isophote map of the face-on spiral NGC 628 (Shostak \&\
van der Kruit 1984). The isophote interval is 0.5 mag. The last three
isophotes show a substantially smaller average spacing than the ones
at higher surface brightness.}
\end{figure}

Evidently, the stellar
material moving near these cutoffs constitutes that with the highest
specific angular momentum in the
disk and it is not unreasonable to expect that the distribution
of specific angular momentum should have a reasonably well-defined upper
limit. Mestel (1963) has shown that the distribution of
specific angular momentum in disks of spiral galaxies closely resembles
that of a uniformly rotating, uniform density
sphere. 

If $h_{\rm s}$ is the specific angular momentum and $M(h_{\rm s})/M$ the
fraction of the mass with specific angular momentum less than or equal
to $h_{\rm s}$, then this distribution for the uniformly rotating, 
uniform density ``Mestel'' sphere is
$$
{M(h_{\rm s}) \over M} = 1 - \left(1 - {h_{\rm s} \over 
h_{\rm max}}\right)^{3/2},
$$
where $h_{\rm max}$ is the maximum specific angular momentum at the
``equator'' of the surface of the sphere. 
If this sets in a flat disk with a flat
rotation curve with rotation velocity $V_{\rm m}$ with detailed conservation
of angular momentum (Fall \&\ Efstathiou 1980), then
the  resulting surface density distribution is close to an exponential 
(Gunn 1982) and van der Kruit (1987) showed that the scalelength $h_{\rm R}$ of 
this distribution is
$$
h_{\rm R} \sim {h_{\rm max} \over {4.5 V_{\rm m}}}.
$$
This implies that the maximum radius of the disk (where the specific angular 
momentum equals $h_{\rm max}$) is $R \sim 4.5 h_{\rm R}$ and a sharp 
truncation occurs there. Crucial in this description is of course that
the disk settles with detailed conservation of angular momentum.

\section{Recent developments.}

Recently, surface photometry in the optical and near-infrared for 
a complete sample of edge-on galaxies has become available in the 
thesis of Richard de Grijs (de Grijs \&\ van der Kruit 1996; de Grijs 1998).
Kregel, van der Kruit \&\ de Grijs (in preparation; see also the poster by
Kregel \&\ van der Kruit in this volume) are in the process 
of re-analysing these data. At this time only the disk scale parameters
have been rederived. De Grijs, Kregel \&\ Wesson (2000) have done 
a first analysis looking for possible truncations in four of these systems.
They find that {\it all four} have truncations in their disks and at least 
three are very symmetric on both sides (see Fig. 3). 

The truncations are not very
sharp; the scalelengths drop to about 2 to 3 kpc. The values for the
truncation radii in terms of the radial scalelengths $R_{\rm
max}/h_{\rm R}$ are 4.3, 3.8, 4.5 and 2.4. There is very little
dependence of the truncation radius itself on color (in
$B$, $V$ and $I$). However, scalelengths are known to
vary with wavelength (de Jong 1996a,b) and the ratio $R_{\rm
max}/h_{\rm R}$ then remains color dependent.

Pohlen, Dettmar \&\ L\"utticke (2000) and Pohlen et al. (2000) 
have recently analysed a sample of 31 edge-on
galaxies. They fitted three-dimensional one-component models to the
observed surface brightness distributions. Then {\it all} have sharp
truncations (their algorithm always fits a truncation to
the data, although in principle its radius could become infinite). These
occur at $R_{\rm max} = 2.9 \pm 0.7 \;\; h_{\rm R}$, significantly less
then 4.5. They argue that dust absorption could raise this value by 0.5
at most. Although this ratio does not correlate with Hubble type, it
does become on average smaller with increasing scalelength. The Kregel 
et al. and the Pohlen et al. samples have three galaxies in common. 
For two the determined scalelengths agree at a satisfactory level.
\begin{figure}
\plotone{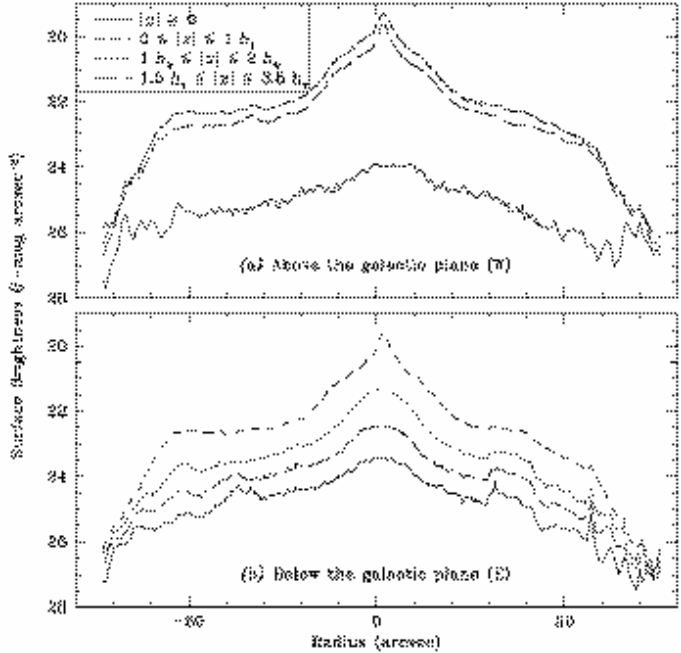}
\caption{Vertically averaged radial surface brightness profiles of ESO
416-G25 at various $z$-heights (de Grijs, Kregel \&\ Wesson 2000).}
\end{figure}

The sample of Pohlen et al. is not complete in a statistical sense;
rather it is weighted heavily in favour of {\it large} scalelengths. 
In Table 1 I compare the distribution of $R_{\rm max}/h_{\rm R}$
in the three available samples of edge-on galaxies.
\begin{table}
\caption{The distribution of the ratio of
truncation radius to disk scalelength in the three samples of
edge-on galaxies}
\begin{tabular}{|c|c c|c c|c c||}
\hline
$h_{\rm R}$ &\multicolumn{2}{c|}{v.d.Kruit \&\ Searle }
&\multicolumn{2}{c|}{de Grijs et al.} & \multicolumn{2}{c||}{Pohlen
et al. } \\
(kpc) & $n$ & $R_{\rm max}/h_{\rm R}$ &
$n$ & $R_{\rm max}/h_{\rm R}$ & $n$ & $R_{\rm max}/h_{\rm R}$ \\
\hline
0 - 6 & 7 & $4.2 \pm 0.5$ & -- & -- & 10 & $3.3 \pm 0.7$ \\
6 - 10 & -- & -- & 2 & 4.3, 2.4 & 12 & $3.1 \pm 0.5$ \\
10 - 15 & -- & -- & 2 & 3.9, 4.5 & 7 & $2.2 \pm 0.5$ \\
\hline
\end{tabular}
\end{table}

The results of de Grijs et al. and of Pohlen et al. for
a large part concern galaxies with very large disk scalelengths (even
larger than in our Galaxy or M31). From de Jong (1996) we can estimate
that in a volume-complete sample of disk galaxies somewhat less than
1\%\ has a scalelength larger than 6 kpc. 

So we may conclude:\\
$\bullet $ Truncations indeed occur in many stellar disks, are often symmetric
and seem not dependent upon color.\\
$\bullet $ The ratio $R_{\rm max}/h_{\rm R}$ appears often less then
4.5. This would imply that truncations should be more easily
observable in moderately inclined systems.\\
$\bullet $ Current samples may not be representative and very strongly
biased towards disks with the largest scalelengths.\\
$\bullet $ The value for the scalelength is crucial (and
color-dependent) and it is important to compare the fitting techniques 
that are being used.

\begin{figure}
\plotone{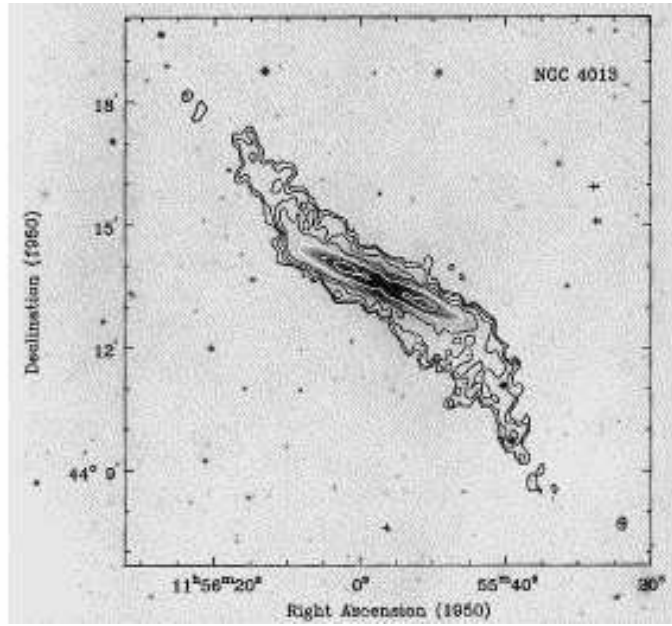}
\caption{The HI-distribution in NGC 4013 (Bottema 1996). The warp sets in
quite suddenly at the sharp edge of the optical disk. 
}
\end{figure}

The situation in our Galaxy is at present unclear. On the basis of the
distribution of OB-stars and HII-regions, one would expect the truncation
radius to be 20--25 kpc. 
The value for the disk scalelength in the Galaxy is 
still under discussion. Recent analyses of
the near-IR and COBE data seem to suggest a disk scalelength of order 2.5
kpc (e.g. Freudenreich 1996, 1998), while the truncation radius of the 
disk would occur at about 12 kpc. The latter seems at variance with the
occurence of HII-regions at much larger radii. Also note that for a
galaxy with a rotation velocity of about 220 km/s, the expected value
for the disk scalelength would be about 4--5 kpc. 
For a fuller and more detailed discussion see van der Kruit (2000).

\begin{figure}
\plotone{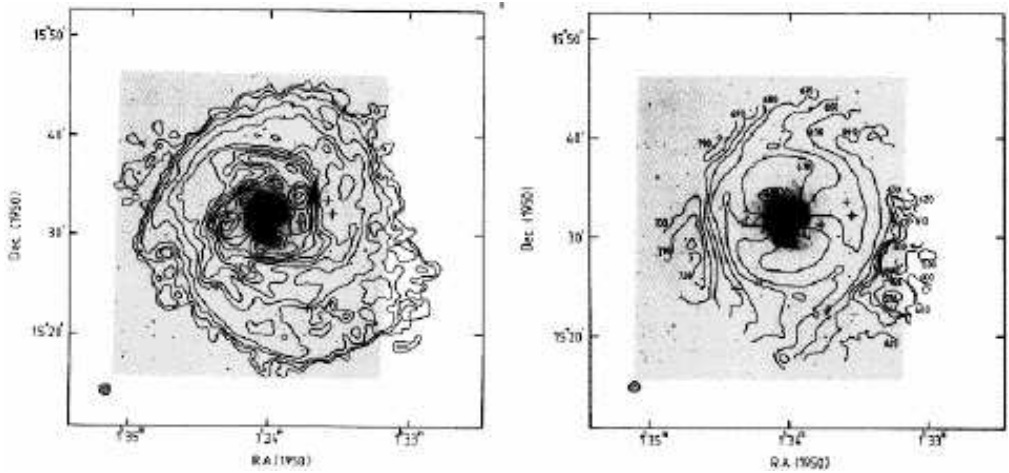}
\caption{The HI distribution (left) and velocity field of the face-on 
spiral NGC 628 (Kamphuis \&\ Briggs 1992). At about the edge of
the stellar disk the velocity field suddenly shows deviations from the
rotation pattern in the inner parts; the plane of the HI layer goes through
that of the sky in a definite warp. Note that the HI surface density
shows the continuation of the spiral pattern through the onset of the
warp.
}
\end{figure}

\section{Beyond the edge}

Various studies (S\'anchez-Saavreda et al. 1990; Florido et al. 1991, de
Grijs 1997) have indicated that warps exist in most stellar
disks. However, these constitute a maximum excursion from
the plane of the disk of a only few percent.

Warps in the HI layer are much larger and extend much further out. The 
archetypical example is NGC 4013 (Bottema, Shostak, \&\ van der Kruit
1987; Bottema 1995,1996). 
The HI warp starts at about the truncation of the stellar disk
and is accompanied by a significant drop in rotation velocity. The
latter suggests a truncation in the disk mass distribution as well as in
the light. Also the HI surface density drops at the start of the warp
and abruptly flattens off. 

A face-on spiral with an HI-warp is NGC 628 (Shostak \&\ van der Kruit
1984; Kamphuis \&\ Briggs 1992). The velocity field suggest a warp in
the HI, starting at the edge of the optical disk. There is no clear 
feature in the rotation curve there
(but it might be difficult to observe at NGC 628's small inclination), 
but again the HI surface density
distribution suddenly flattens off. Note that also the spiral structure
in the gas is continuing right through the onset of the warp. Kamphuis
and Briggs argue for a recent accretion event as the cause of the warp
on the basis of the motions in the gas becoming more chaotic with 
increasing galactocentric radius.

Ferguson et al. (1998) find evidence for faint HII-regions in the
extreme outer parts of three spiral galaxies, among which NGC 628. In the
latter H$\alpha $ is observed out to twice the optical radius. So,
some star formation seems to be going on there. Interestingly,
also a sharp drop is seen in the azimuthally averaged H$\alpha $ surface
brightness exactly at the edge of the disk. Abundance determinations in
these extreme outer
regions indicate values of order 10 to 15\%\ of solar in O/H and 
20 to 25\%\ in N/O abundance. Also, the Balmer decrements provide clear
evidence that the internal extinction in these outer parts is very low
($A_{V} \sim 0-0.2$ mag), indicating dinimished dust contents. So,
outer disks appear relatively unevolved compared to inner disks.

I have discussed only two clear examples here. However, many of these
features are found also in other galaxies.
I conclude:\\
$\bullet $ Stellar disks are usually warped, but only moderately so.\\
$\bullet $ Many spiral galaxies have HI warps and these
generally start near the truncation radius of the
stellar disk. The HI surface surface density suddenly becomes much
flatter with radius.\\
$\bullet $ In some galaxies (notably NGC 4013, 891, 5907) there is a
drop in the rotation curve at the edges of the stellar disks.\\
$\bullet $ Some star formation goes on in the extreme outer regions,
but the heavy element abundance and dust content are very low.\\
$\bullet $ All evidence is consistent with the notion that the outer
gaseous parts of the disks constitute recently accreted material,at least
accreted after the formation of what is now the stellar thin disk.

\section{The origin of the truncations}

The origin of the truncations in stellar disks is still unclear.
Originally Fall \&\ Efstathiou (1980) and van der Kruit \&\ 
Searle (1982) suggested, that the edges
correspond to the positions where differential rotation becomes 
able to stabilize the gas layer (according to the Goldreich \&\ Lynden-Bell 
[1965] criterion), so that star formation is prohibited
beyond that radius.
Kennicutt (1989) and others have argued that
the truncation radius is the position where the gas density drops below
a critical value for star formation. 
This is then regulated by the Toomre (1964) stability parameter $Q$. 
These two hypotheses are
not made compatible easily with the sudden drop in the rotation curves
at $R_{\rm max}$.

The notion that the truncation radius results from the
maximum specific angular momentum present in the material from which
the (presently stellar) disks formed is in itself straightforward. The
paradigm, where the initial material resembles a Mestel sphere
with uniform density and uniform rotation, and where the collapse into
a disk occurs with detailed conservation of angular momentum (even if 
it occurs at a slow rate), provides a good explanation for the exponential
nature of the disk surface brightness (and density). However, it also
predicts a definite position for the truncation at
$R_{\rm max}/h_{\rm R} \sim 4.5$, and that seems higher
than is observed. This model would require therefore some redistribution of
angular momentum.

Another possibility is very slow disk formation, where
the truncation radius would then be the extent out to which the disk has 
presently formed. This and models of viscous processes regulating star
formation (e.g. Lin \&\  Pringle 1987; Yoshii \&\  Sommer-Larsen 1989)
predict no particular value for the truncation radius in terms of the disk
scalelength. Tidal interactions are only a possibility in
some systems; truncations exist in galaxies independent of their
environment.

\end{document}